\documentclass[aps,twocolumn]{revtex4}
\usepackage{graphics} \begin{document}
\newcommand{\bstfile}{aps} %alternative styles: osa, prasty or revtex
\draft
\title{Modulated conjugation as a means for attaining a
record high intrinsic hyperpolarizability}

\author{Javier P\'erez-Moreno}

\address{Department of Chemistry, University of Leuven, Celestijnenlaan 200D, B-3001 Leuven, Belgium}
\email{Javier.PerezMoreno@fys.kuleuven.be}

\author{Yuxia Zhao}

\address{Technical Institute of Physics and Chemistry, Chinese Academy of Sciences, 100101, Beijing, China}
\email{yuxia.zhao@mail.ipc.ac.cn}

\author{Koen Clays}

\address{Department of Chemistry, University of Leuven, Celestijnenlaan 200D, B-3001 Leuven, Belgium}
\email{Koen.Clays@fys.kuleuven.be}

\author{Mark G. Kuzyk}

\address{Department of Physics and Astronomy, Washington State University, Pullman, Washington 99164-2814}
\email{kuz@wsu.edu}

\date{\today}

\begin{abstract}
We report on a series of chromophores that have been synthesized
with a modulated conjugation path between donor and acceptor.
Hyper-Rayleigh scattering measurements of the best molecule show an
enhanced intrinsic hyperpolarizability that breaches the apparent limit of all previously-studied molecules.
\end{abstract}

\maketitle

\vspace{1em}

Over the last 3 decades, many novel molecules have been designed and
synthesized to improve the nonlinear response for a variety of
applications. Quantum calculations using sum rules have been used to
place an upper-bound on the molecular susceptibilities;
\cite{kuzOpt2000,kuzPRL,kuzOL2003,kuzyk03.02} but, the largest
nonlinear susceptibilities of the best molecules fall short of the
fundamental limit by a factor of
$10^{3/2}$.\cite{kuzyk03.02,kuzyk03.05}  A thorough analysis shows
that there is no reason why the molecular hyperpolarizability can
not exceed this apparent limit.\cite{Tripa04.01} In this letter, we
report on a novel set of molecules where the one with modulated
conjugation\cite{zhou06.01a} is found to have a hyperpolarizability
that breaches the apparent limit of all previously-measured molecules.

Past work has shown that the polarizability is largest when the
potential energy function oscillates in a way that localizes the
eigenfunctions on different parts of the molecule.\cite{zhou06.01a}
This type of oscillation can be designed into a bridge that separates
the donor and acceptor ends of a chromophore by varying the degree
of conjugation.  Our approach is based on the well-known difference in aromatic stabilization energy between benzene and heterpentacyclics, such as thiophene rings.\cite{refa}

\begin{figure}
\includegraphics{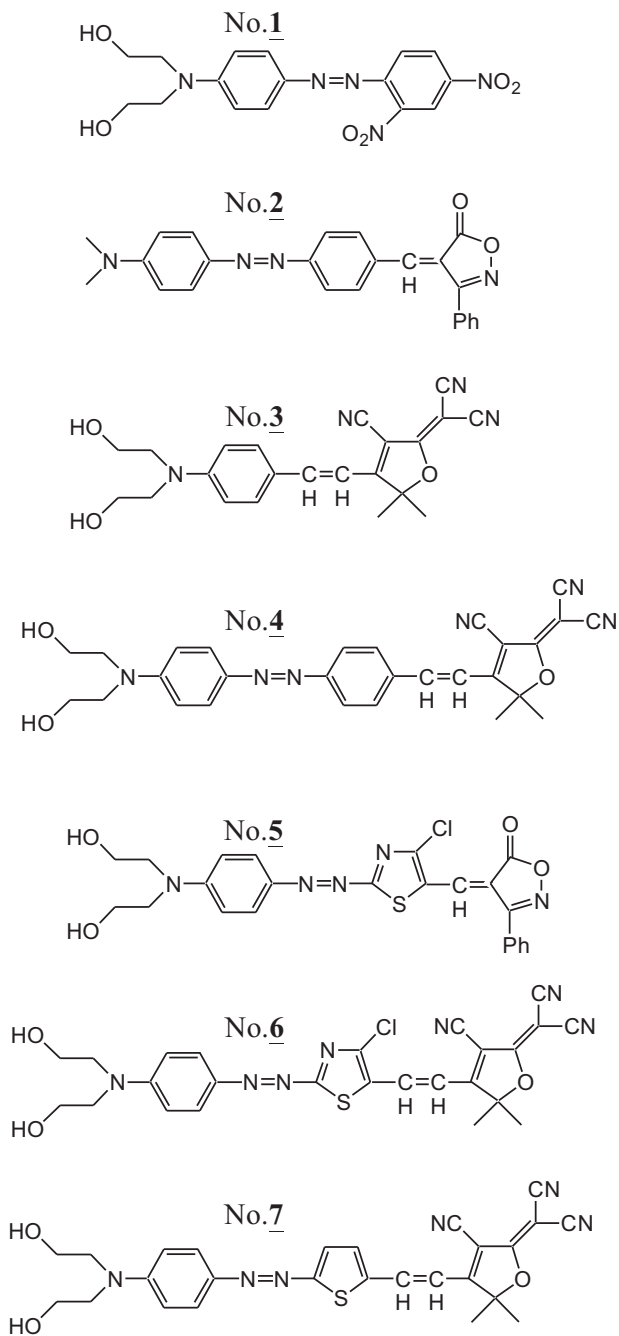}
\caption{Molecules.} \label{fig:molecules}
\end{figure}

Figure \ref{fig:molecules} shows the series of molecules under
study.  The synthesis and the details of the linear and nonlinear optical characterization of this series of compounds will be published elsewhere.\cite{refb}

\begin{table}\caption{Molecular Properties. Uncertainty in hyperpolarizability measurements is about 10\%.\label{data:table}}
\begin{tabular}{ccccc}
\hline Molecule & $\lambda_{MAX}$ & $N$ & $\beta_0$ & $\beta_0 /
\beta_{MAX}$ \\
 & (nm) & & $10^{-30} \, (esu)$ & \\
\hline \hline
1  & 551 & 18 & 110 & 0.0208 \\
2  & 540 & 20 & 110 & 0.0190 \\
3  & 602 & 18 & 240 & 0.0332 \\
4  & 567 & 26 & 340 & 0.0334 \\
5  & 695 & 18 & 280 & 0.0234 \\
6  & 691 & 24 & 735 & 0.0408 \\
7  & 677 & 24 & 800 & 0.0477 \\
\hline
\end{tabular}
\end{table}

\begin{figure}
\includegraphics{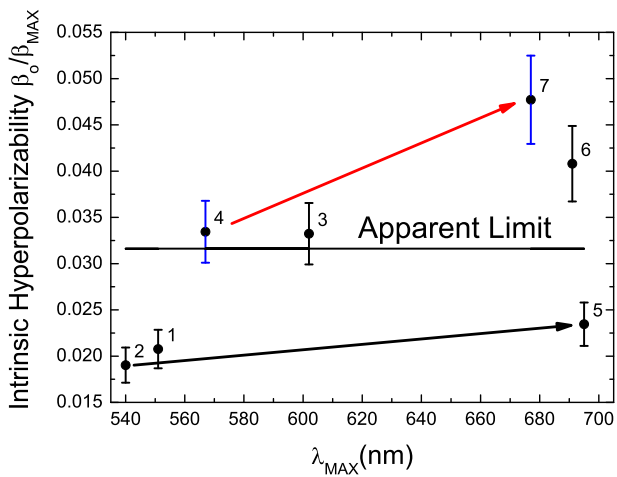}
\caption{Zero-frequency hyperpolarizability normalized to the
fundamental limit, as a function of wavelength of maximum
absorption.} \label{fig:data}
\end{figure}

The hyperpolarizability, $\beta$, was determined at $800 \, nm$
using Hyper-Rayleigh scattering.  The zero-frequency
hyperpolarizability, $\beta_0$ was determined using the two-level
model.  Table \ref{data:table} shows the measured molecular
properties and Figure \ref{fig:data} shows a plot of $\beta_0$,
normalized to the fundamental limit of the hyperpolarizability (this ratio is called the intrinsic hyperpolarizability, which is scale-invariant),
where the fundamental limit, $\beta_{MAX}$, is given by
\begin{equation}
\beta_{MAX} = \sqrt[4]{3} \left( \frac {e \hbar} {\sqrt{m}}
\right)^3 \cdot \frac {N^{3/2}} {E_{10}^{7/2}} ,
\end{equation}
where $N$ is the number of electrons ($N$ is determined by counting methods described in the literature\cite{jchemphys03Kuzyk,jchemphys05xavi}), $E_{10}$ the energy difference between state $1$ and $0$, $e$ is the electron charge, $\hbar$ Planck's constant and $m$ the mass of the electron.
The horizontal line in Figure \ref{fig:data} represents the apparent limit defined by the best past measurements, which is a factor
of $10^{3/2}$ below the fundamental limit.

No single molecule has ever been reported to breach the apparent
limit, though some have come close.  For example, May and coworkers
have shown the the second hyperpolarizability gets within a factor
of 2 of the apparent limit.\cite{May05.01}  Wang and coworkers, on
the other hand, have reported breaking through the apparent
limit.\cite{wang04.01}  However, a close analysis shows that their
chromophores, being part of a cross-linked system,  were strongly
interacting -- leading to an under-counting of the number of
electrons.  So, we believe our reported values to be the first
example of single chromophores that breach the limit with record intrinsic hyperpolarizability.

Our molecular design focuses on modulating the amount of aromatic stabilization energy along the conjugated bridge between the donor and the acceptor. To induce the desired modulation, aromatic moieties with a different degree of aromaticity make up the asymmetrically substituted $\pi$-bridge. As an example, molecules 4 and 7 are both azo dyes, but while molecule 4 has 2 benzene moieties with identical (36 kcal/mol, or 1.57 eV) aromatic stabilization energy, molecule 7 has a benzene and a thiophene moiety. The latter is well known to have a reduced aromatic stabilization energy (29 kcal/mol, or 1.25 eV). This results in a significant variation of the degree of aromaticity for molecule 7, or in a modulation of the conjugation between the donor and acceptor. This degree of
conjugation modulation yields an enhancement in the
hyperpolarizability that breaches the apparent limit well outside the
range of experimental uncertainty of 10\%.  Note that a
demodulation technique is used to eliminate the
background contribution from two-photon fluorescence,\cite{clays91.01,clays92.01,10new} insuring that the measured values are not overestimated due to this known source of systematic error.

Molecule 6 is isoelectronic to molecule 7, but due to the additional Cl atom in molecule 6, steric hindrance induces a twist in the conjugation path, resulting in a decreased hyperpolarizability.  Molecules 2 and 5 are homologues of 4 and 7; and show a similar
enhancement.  However, the larger and more geometrically linear
molecules show a more dramatic effect, which is predicted by the
theory (steric hindrance caused by the chlorine atom suppresses the enhancement of molecule 5.).\cite{zhou06.01a}  In particular, the best molecules are ones
that are long with many undulations in the potential energy function, which
allows for the electron densities of the eigenstates to be well
separated.  So, future design strategies should focus on longer
molecules with stronger modulation of conjugation.

In addition to increased length, future efforts
must also focus on keeping the chain linear.  Special attention should be devoted to mimic the optimal undulation\cite{zhou06.01a} by making use of not only benzene and thiophene, but also of other aromatic moieties that exhibit an even wider range of stabilization energies (like pyrrole and furan with aromatic stabilization energy values of 22 and 16 kcal/mol, or 0.98 and 0.69 eV, respectively).

While our best measured values of the hyperpolarizability are still
more than an order of magnitude from the fundamental limit, our
design strategy appears to be a promising new paradigm for making
better molecules.

{\bf Acknowledgements: } JPM acknowledges a PhD fellowship from the University of Leuven.  KC acknowledges support from the Belgian government (IUAP/5/3), from the Fund for Scientific Research in Flanders (G.0297.04) and from the Research Council of the University of Leuven.  MGK thanks the National Science Foundation (ECS-0354736) and Wright Paterson Air Force Base for generously supporting this work.

%\bibliography{\bibs}

%\clearpage

\end{document}